\documentclass[preprint, aps, pre, preprintnumbers,amsmath,amssymb,showpacs]{revtex4}

\usepackage[final]{graphicx}  

\newcounter{saveeqn}

\begin{document}
\title{
Wave number dependence of the transitions between travelling and standing vortex waves
and their mixed states in the Taylor-Couette system}
\author{A.~Pinter\footnote[1]{Electronic address: kontakt@alexander-pinter.de}, M.~L\"ucke, and Ch.~Hoffmann}
\affiliation{Institut f\"ur Theoretische Physik, Universit\"at des Saarlandes, Postfach 151150, D-66041 Saarbr\"ucken, Germany}
\date{\today}

\begin{abstract}
Previous numerical investigations of the stability and bifurcation properties
of different nonlinear combination structures of spiral vortices in a counter-rotating 
Taylor-Couette system that were done for fixed axial wavelengths are
supplemented by exploring the dependence of the vortex phenomena waves on
their wavelength. This yields information about the experimental and numerical
accessability of the various bifurcation scenarios. Also backwards bifurcating
standing waves with oscillating amplitudes of the constituent traveling waves
are found.
\end{abstract}



\maketitle


Recently the stability exchange between travelling waves (TWs)
\cite{abbreviations} and standing waves (SWs) of spiral vortices in the
Taylor-Couette system has been investigated by full numerical simulations and
a coupled amplitude equation approximation \cite{PLH06}. TWs and SWs have a
common onset as a result of a primary, symmetry degenerate oscillatory
bifurcation. The SW solution is a nonlinear superposition of mirror symmetric,
oppositely propagating TWs with equal amplitudes. At onset either the TW or
the SW solution is stable \cite{GS87,DI84}. Then, at larger driving there is a
secondary bifurcation that leads to a stability exchange between the two
solutions. This exchange is mediated by mixed patterns that establish in
solution space a connection between a pure TW and a pure SW. The mixed
structures consist of a superposition of oppositely propagating TWs with {\em
temporally constant, non-equal} amplitudes. The TWs investigated in
\cite{PLH06} are initially stable while the SWs gain stability later on.

There is a second variety of mixed states in which the TW amplitudes {\em
oscillate in time} in counterphase. This stable solution bifurcates out of the
SW at even higher driving rates via a Hopf bifurcation \cite{PLH08} in which
the aforementioned SWs lose their stability. These results have been found by
full numerical simulations of the vortex flow in a Taylor-Couette system
\cite{CI94,T94} with counter-rotating cylinders of radius ratio $\eta=0.5$
with methods described in \cite{HLP04}. The calculations were done for a fixed
axial wavelength $\lambda$ by imposing axially periodic boundary conditions
\cite{end-effects}.

Here we investigate and show how stability, bifurcation properties, and the
spatiotemporal behavior of the aforementioned structures change with
$\lambda$. Thus, this report provides information for
future simulations and experiments with finite length set-ups and, say,
nonrotating lids that close the annular gap between the cylinders at the ends:
Since the height of the system influences the wavelength of the vortex
structures and with it their properties the prior knowledge of their
$\lambda$-dependence is of significant interest.

{\it Structures} ---
The following structures have been investigated:
({\em i}) Forward bifurcating
TWs consisting of left handed spiral vortices (L-SPI) or of right
handed spiral vortices (R-SPI) that are mirror images of each other.
L-SPI (R-SPI) travel in the annulus between the two cylinders axially into
(opposite to) the direction of the rotation frequency vector of the
inner one, i.e., in our notation upwards (downwards) \cite{HLP04}. ({\em ii})
Forward bifurcating SWs that consist of an equal-amplitude nonlinear
combination of L-SPI and R-SPI. These SWs are called ribbons (RIBs) in the
Taylor-Couette literature \cite{TESM89,LPA03}. ({\em iii}) So-called
cross-spirals (CR-SPI), i.e., combinations of L-SPI and R-SPI with
different stationary
amplitudes. They provide a stability transferring connection between
TW and SW solution branches \cite{CI94,PLH06}. And, finally, ({\em iv})
oscillating cross-spirals (O-CR-SPI). Therein, the amplitudes of the
TW constituents of the SW, i.e., the amplitudes of
L-SPI and the R-SPI oscillate in counterphase around a common mean
\cite{PLH08}.
The vortex structures ({\em i})-({\em iv}) are axially and azimuthally
periodic with axial wave number $k=2\pi/\lambda$ and azimuthal wave number
$M =2$ in our case.

{\it Control- and order parameters} ---
The control parameters are the Reynolds numbers $R_1>0$ and $R_2<0$ defined by
the rotational velocities of the inner and outer cylinder, respectively.
As order parameters we use the amplitudes
\begin{equation} \label{EQ:u2pm1}
A(t)=u_{2,1}(t), \quad  \quad B(t)=u_{2,-1}(t)\,
\end{equation}
of the dominant critical modes of the radial velocity $u$ at mid-gap in the
double Fourier decomposition in azimuthal and axial direction. In
Eq.~(\ref{EQ:u2pm1}) the indices
$m=2$ and $n=\pm 1$ identify azimuthal and axial modes, respectively. Note
that for SPI, CR-SPI, and RIB structures investigated here the moduli
in Eq.~(\ref{EQ:u2pm1}) are constant. On the other hand, in O-CR-SPI the
moduli $|A(t)|$ and $|B(t)|$ oscillate in counterphase around a common mean.
Therefor we use the difference of the squared moduli
$D(t)=(|A(t)|^2-|B(t)|^2)/2$ and its oscillation amplitude $\widetilde D$ to
describe the bifurcation from the RIB solution $(\widetilde D=0)$ to O-CR-SPI
$(\widetilde D \neq 0)$.

\begin{figure}
\includegraphics[clip=true,width=8.6cm,angle=0]{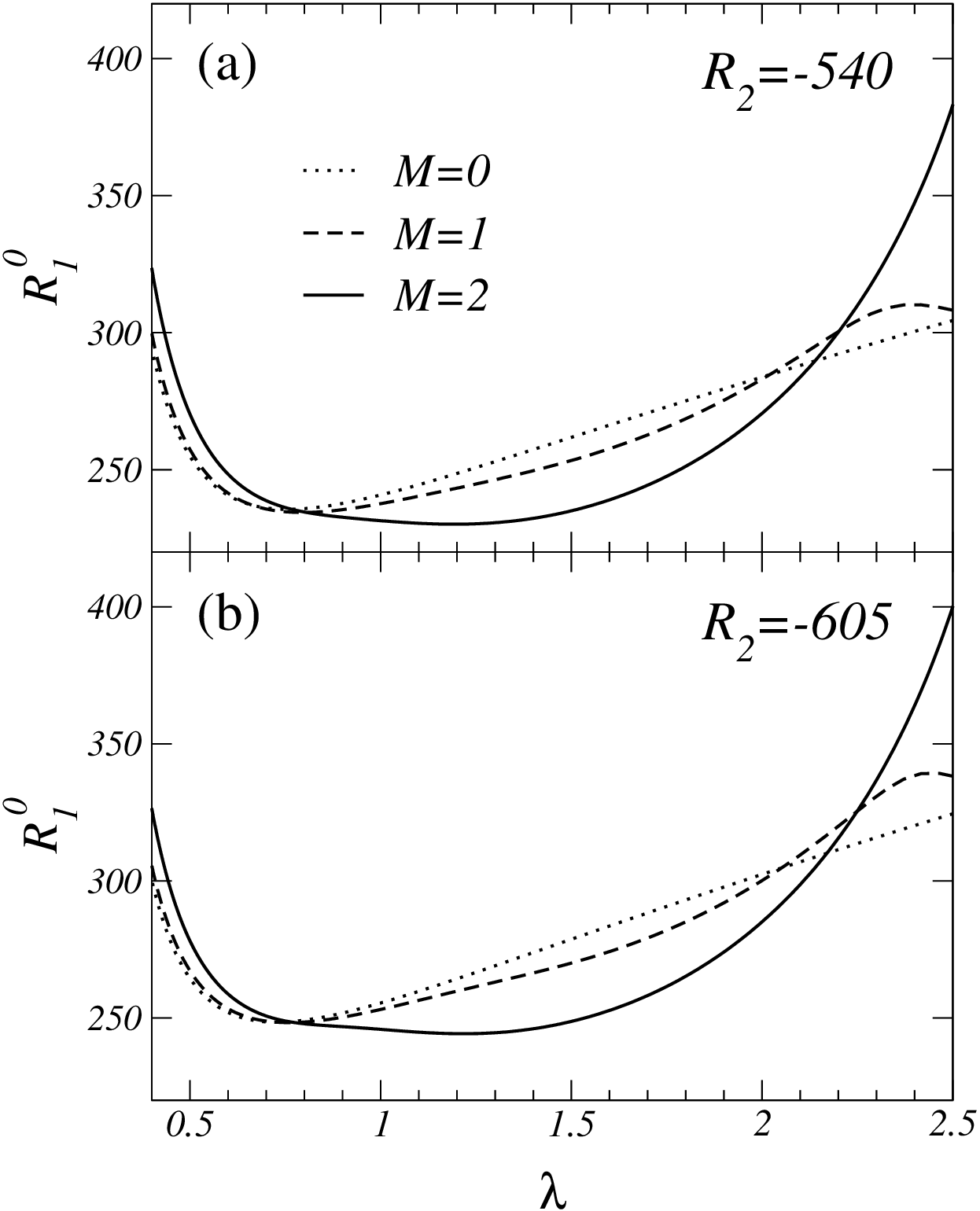}
\caption{Bifurcation thresholds $R_1^0$ of $M=0$ Taylor vortex flow and of SPI
and RIB with azimuthal wave numbers $M=1$ and $M=2$ versus axial wavelength
$\lambda$ for
different $R_2$ as indicated. \\
\label{FIG1}}
\end{figure}

{\it  $\lambda$-dependence of the bifurcation scenario ---}
Fig. \ref{FIG1} shows the $\lambda$-dependence of the bifurcation thresholds
$R_1^0$ for \mbox{$M=1$} and \mbox{$M=2$} SPI and RIB and $M=0$ Taylor vortex
flow. These results were obtained from a linear stability analysis
\cite{PLH03} of the basic circular Couette flow. For the two characteristic
values $R_2=-540$ and $-605$ shown there the $M=2$ SPI and RIB have the lowest
threshold for a wide range of $0.8 \lesssim \lambda \lesssim 2.1$.

\begin{figure}
\includegraphics[clip=true,width=8.6cm,angle=0]{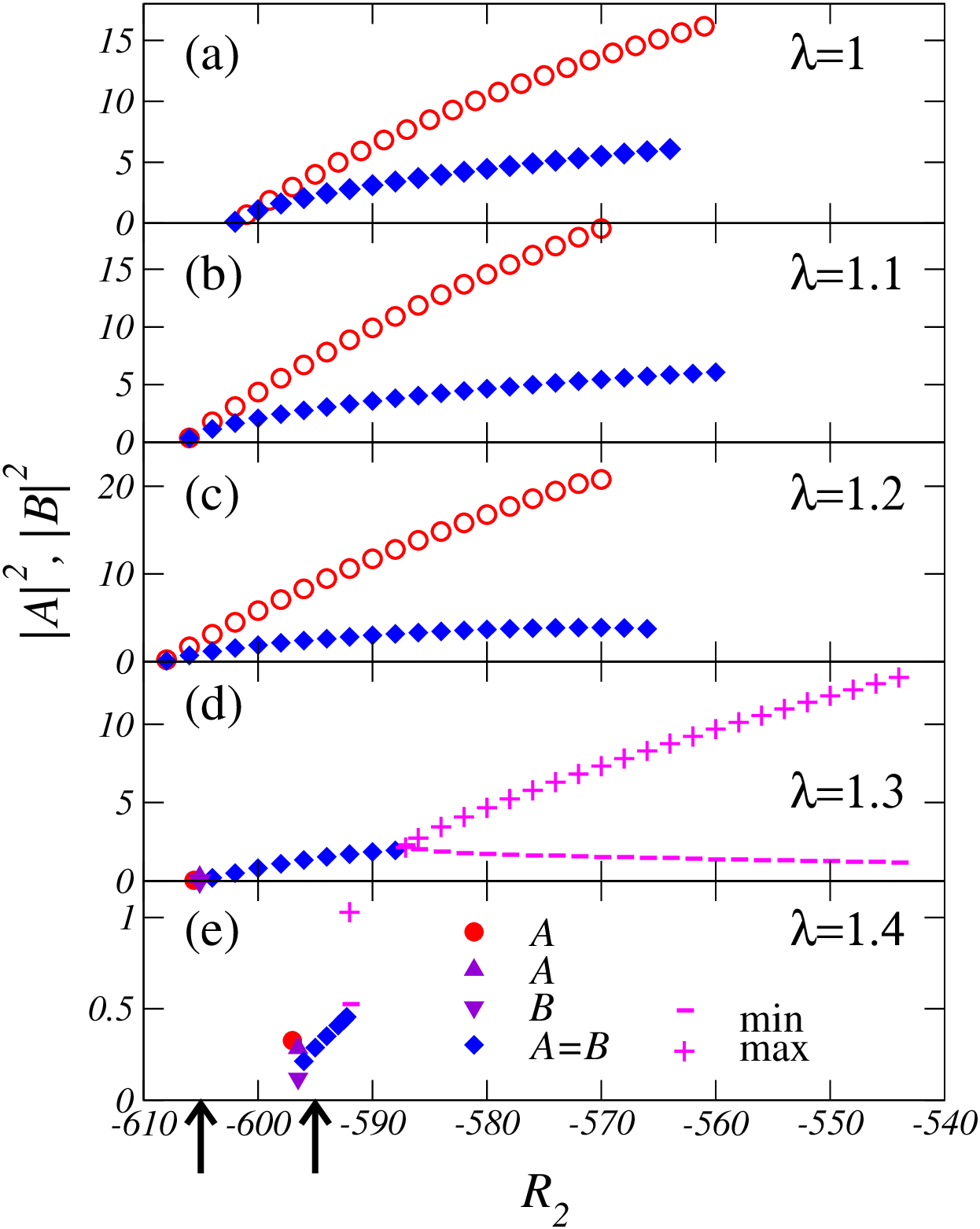}
\caption{(Color online). Influence of the axial wavelength $\lambda$ on the bifurcation
behavior for fixed $R_1=240$. Solid (open) symbols denote stable (unstable)
structures. Symbols show \mbox{L-SPI} (red circles,
$A\neq0, B=0$), L-CR-SPI (violet triangles, $|A|>|B|$), RIB (blue diamonds,
$A=B$), and stable \mbox{O-CR-SPI} (magenta). In the latter $|A(t)|$ and
$|B(t)|$ oscillate between the $+$ and $-$ symbols. In (d) and (e) only stable
structures are shown. The arrows refer to the $R_2$ values of Fig. \ref{FIG3}.
\label{FIG2}}
\end{figure}

In Fig. \ref{FIG2} we show for different $\lambda$ bifurcation diagrams as
functions of $R_2$ for a fixed $R_1=240$. For $1 \lesssim \lambda \lesssim
1.2$, figure parts (a)-(c), the bifurcation properties are quite similar: SPI
and RIB bifurcate supercritically out of the circular Couette flow, SPI (RIB)
are unstable (stable) at onset, and there is no stability exchange
in the range of $R_2$ of Fig. \ref{FIG2}.

By contrast, for $\lambda=1.3$ and $1.4$, in (d) and (e),
respectively, there are different interesting stability exchanges.
Here, L-SPI $(A \neq 0, B=0)$ and R-SPI $(A =0, B\neq0)$ are initially
stable while the RIB state $(A=B)$ is initially unstable. But RIB gain
stability almost immediately thereafter: the stability transfer from L-SPI
or R-SPI to RIB is
mediated within a very small interval by the L-CR-SPI ($|A|>|B|$) or the
R-CR-SPI ($|B|>|A|$) solution,
respectively \cite{PLH06}. For larger driving, the RIB lose
stability again, when stable oscillating structures, O-CR-SPI, appear via a
Hopf bifurcation \cite{PLH08}. Note, however, that O-CR-SPI
bifurcate forwards for $\lambda=1.3$ but backwards for $\lambda=1.4$, cf.
details further below.

\begin{figure}
\includegraphics[clip=true,width=8.6cm,angle=0]{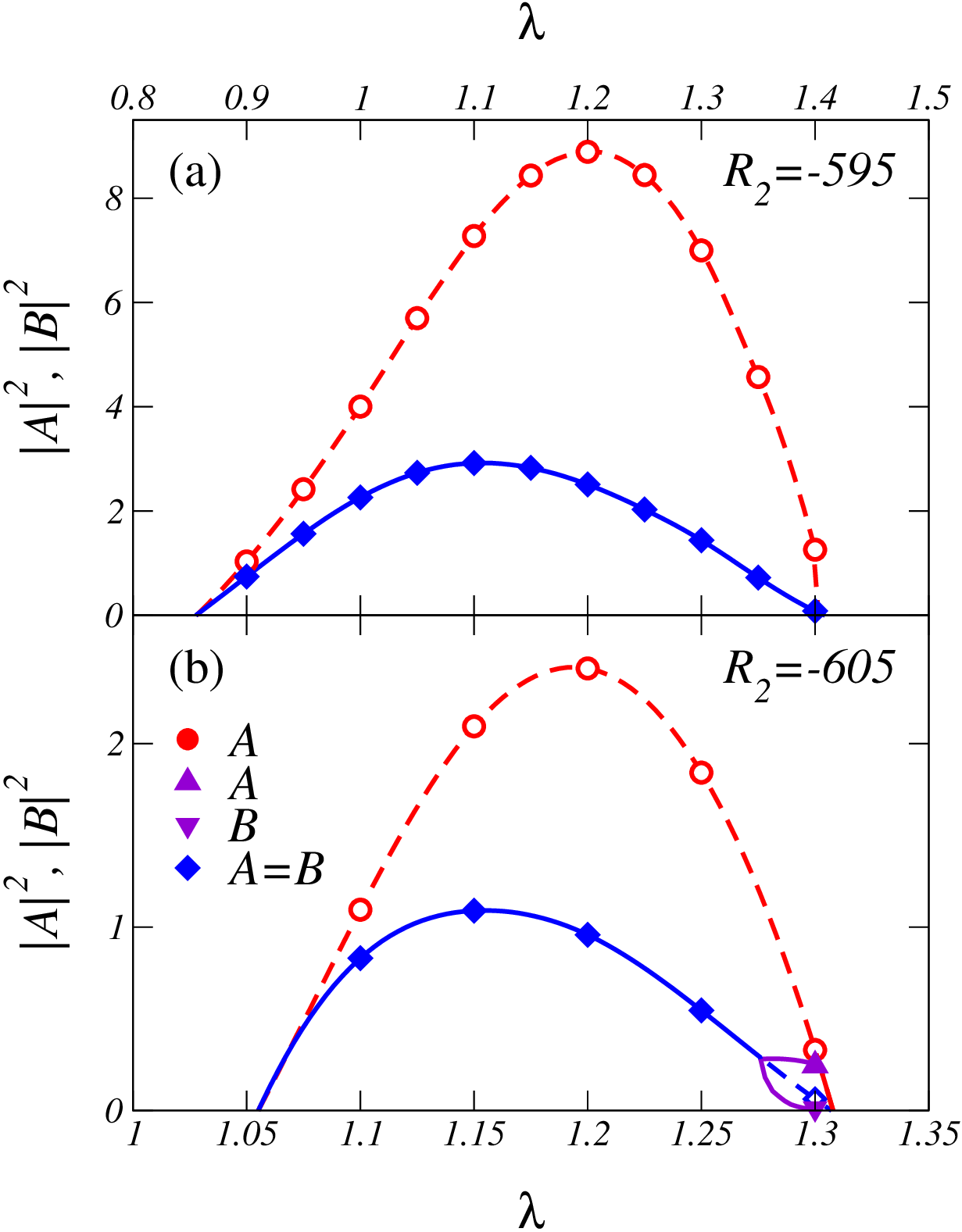}
\caption{(Color online). Bifurcation diagrams of \mbox{L-SPI} (red circles, $A\neq0, B=0$), RIB
(blue diamonds, $A=B$), and L-CR-SPI (violet triangles, $|A|>|B|$) as a function of the axial
wavelength $\lambda$ for fixed $R_1=240$ and the two values of $R_2$ that are
indicated by arrows in Fig. \ref{FIG2}.
Solid (open) symbols denote calculated stable (unstable) structures.
Full (dashed) lines for stable (unstable) solutions branches result from
a spline interpolation. However, the violet L-CR-SPI branches are schematic.
\label{FIG3}}
\end{figure}

In Fig. \ref{FIG3} we show bifurcation diagrams as a function of
$\lambda$ for $R_1=240$ and two different $R_2$ indicated by arrows in
Fig. \ref{FIG2}. In the case of $R_2=-595$ (right arrow in Fig. \ref{FIG2})
SPI are unstable and RIB are stable for all $\lambda$.
For $R_2=-605$ (left arrow in Fig. \ref{FIG2}), on the other hand, this
stability situation --- RIB are stable and SPI are unstable --- applies only
as long as $\lambda \lesssim 1.25$: Then, with increasing $\lambda$, a
stability exchange between RIB and SPI via CR-SPI occurs that is reflected
also at the very left end of the bifurcation diagram in Fig. \ref{FIG2} (d).
So, the interesting stability exchange between TWs and SWs
occurs in a rather narrow wave number band around $\lambda \approx 1.3$.

In figures \ref{FIG2} and \ref{FIG3}, we showed the case where stability is
transferred from L-SPI to RIB. The symmetry degenerated situation where
stability is transferred from R-SPI to RIB via R-CR-SPI is obtained by
exchanging (the symbols for) $A$ and $B$ in these figures.

\begin{figure}
\includegraphics[clip=true,width=8.6cm,angle=0]{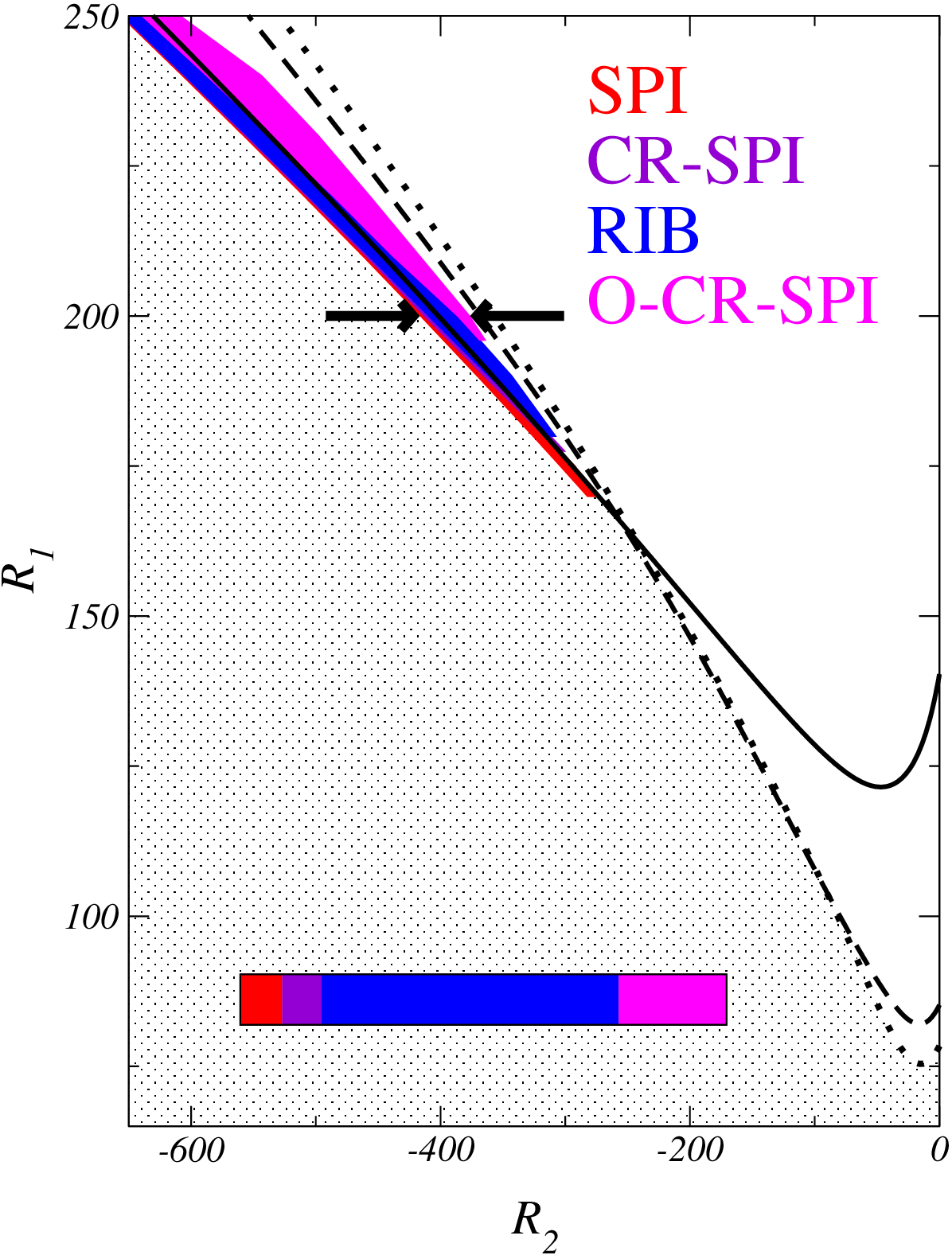}
\caption{(Color). Phasediagram of stable SPI, CR-SPI, RIB, and
O-CR-SPI. In the colored  areas these vortex structures (as indicated by the
color code) with azimuthal wave number $M=2$ and axial wavelength
$\lambda=1.3$ are stable. The white area has not been investigated in this
work. The basic state of circular Couette flow is stable in the dotted region:
Lines are marginal stability boundaries of circular Couette flow against
growth of vortex flow with $\lambda=1.3$ and azimuthal wave numbers $M=2$
(full), $M=1$ (dashed), and $M=0$ (dotted). The bar shows a blow-up of the
region between the black arrows. \label{FIG4}}
\end{figure}

{\it Phase diagram for $\lambda=1.3$ ---}
In view of the above discussed stability exchange process we take a more
detailed look at the case $\lambda=1.3$ for which previous calculations have
been done only at the two Reynolds numbers $R_1=200$ and $240$
\cite{PLH06,PLH07,PLH08}. To that end we provide in Fig. \ref{FIG4} the phase
diagram of the stable, aforementioned $M=2$ vortex structures with fixed
$\lambda=1.3$ in the $R_1-R_2-$ parameter plane. Stable $M=2$ SPI appear first
via a primary forwards bifurcation at the lower left border of the red stripe
in Fig. \ref{FIG4} \cite{bif}. Then, for a fixed $R_1 \gtrsim 190$, we have
observed with increasing $R_2$ always the same stability transfer sequence:
\[
\textrm{SPI} \rightarrow \textrm{CR-SPI} \rightarrow \textrm{RIB} \rightarrow
\textrm{O-CR-SPI}. \]

For lower $R_1$, however, the existence range of stable $M=2$ structures seems
to be more and more confined from above by the appearance of $M=1$ modes at
the respective bifurcation threshold (dashed line): With decreasing $R_1$
first the O-CR-SPI and then the RIB and CR-SPI areas are pinched off
successively.

Note that in all cases the CR-SPI stripe is extremely thin (cf. the blow-up
bar in Fig. \ref{FIG4}) whereas the O-CR-SPI area being quite large should
facilitate a respective experimental observation.

\begin{figure}
\includegraphics[clip=true,width=8.6cm,angle=0]{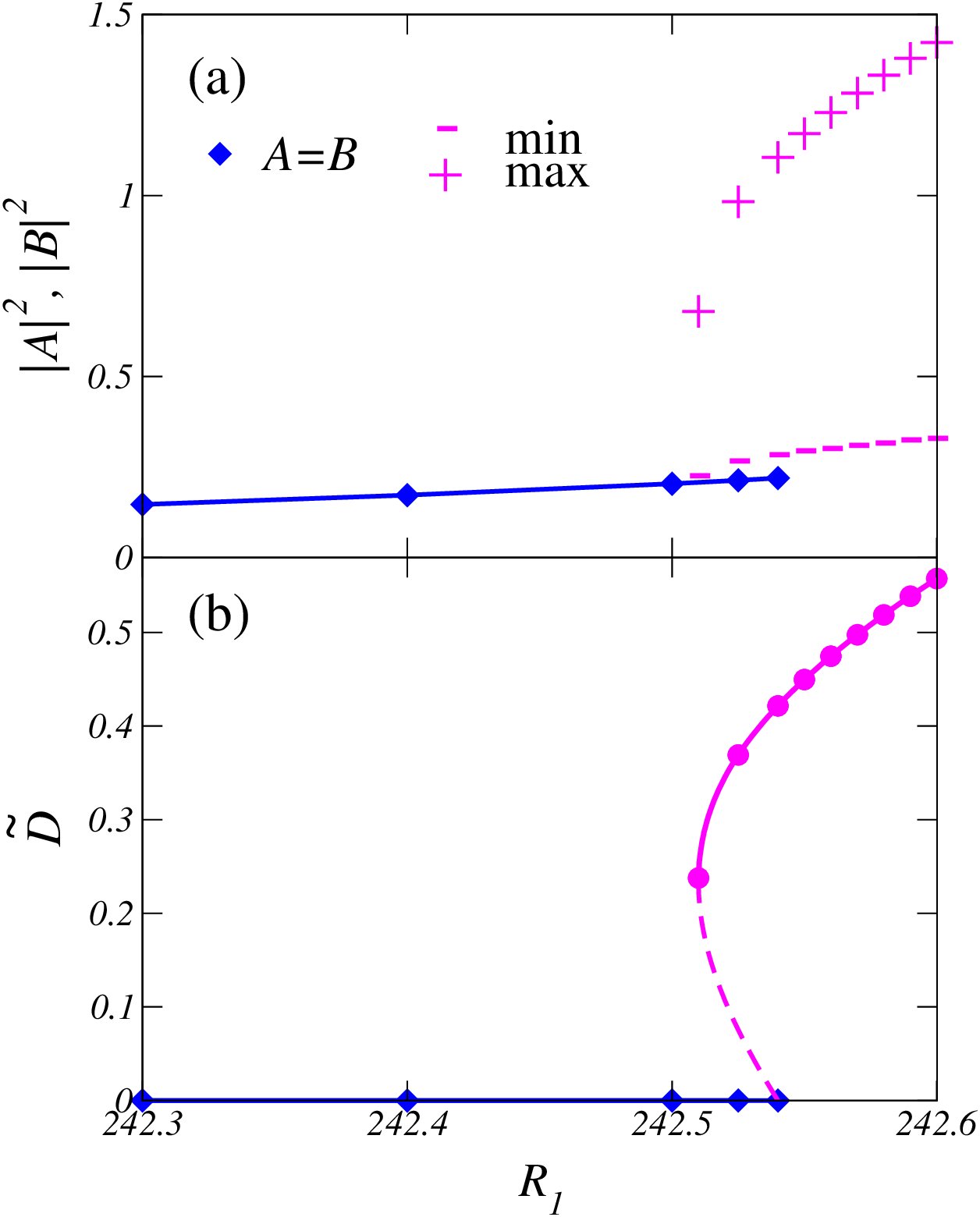}
\caption{(Color online). Backwards bifurcation of \mbox{O-CR-SPI} (magenta circles and $+$ and $-$ signs)
out of RIB (blue squares) for $\lambda=1.4$ and $R_2=-605$ as a function of $R_1$. In (a) squared
amplitudes $|A|^2,|B|^2$ are shown with symbols as explained in
Fig.~\ref{FIG2}. The oscillation amplitude $\widetilde D$ of
$D=\left(|A|^2-|B|^2 \right)/2$ is shown in (b). For RIB \mbox{$\widetilde
D=D=0$}. Full (dashed) lines denoting stable (unstable) solutions were obtained
by spline interpolation. \label{FIG5}}
\end{figure}

{\it Backwards bifurcating O-CR-SPI} ---
As noted already in the discussion related to Fig.~\ref{FIG2}(e), we have
found for the first time backwards bifurcating O-CR-SPI. In Fig.~\ref{FIG5} we
display for fixed $\lambda=1.4$ and $R_2=-605$ as a representative example the
bifurcation properties of this new scenario. Fig.~\ref{FIG5}(a) shows the
squared moduli $|A|^2$ and $|B|^2$ as a function of $R_1$. Diamonds show
stable RIB. They have obtained their stability from the SPI via a CR-SPI
branch connection at smaller $R_1$ outside the plot range of Fig.~\ref{FIG5}.

The $+$ and $-$ signs denote the maximal and minimal amplitudes,
respectively, of the modes $A$ and $B$ that oscillate in counterphase in the
O-CR-SPI. The hysteresis in the transition between stable RIB and stable
O-CR-SPI is best visible with the order parameter $\widetilde D$ in
Fig.~\ref{FIG5}(b).

{\it Conclusion} ---
Our results show that the mixed states of stationary CR-SPI and of O-CR-SPI
should be observable in experimental setups or in finite lengths
numerical simulations when the wavelength of these vortex structures lies in
the interval of
$1.3 \lesssim \lambda \lesssim 1.4$. Therein O-CR-SPI
are stable in a wide range of control parameters. They bifurcate
either forwards or, as we have found here, backwards out of the RIB state
of standing waves. CR-SPI solutions, on the other hand, exist only in a rather
small interval of control parameters.


\newpage

\begin{thebibliography}{999}
\bibitem{abbreviations}
Throughout this manuscript we use the following abbreviations TW: traveling
wave, SW: standing wave, SPI: spiral vortices, L-SPI: left handed spiral
vortices, R-SPI: right handed spiral vortices,  RIB: ribbon, CR-SPI:
cross-spirals, O-CR-SPI: oscillating cross-spirals.
\bibitem{end-effects}
Investigations of traveling spiral vortex waves in finite systems with
non-rotating rigid lids showed that for an aspect ratio of, say, larger than
about 12 the end effects are limited to the lid regions. The dynamics as well
as the structural details of fully developed spiral vortex flow in the bulk
region are widely unaffected by the boundaries -- cf. Figs. 8 and 9 of
\cite{HLP04}. We expect a similar behavior for our case in systems that are
large enough.
\bibitem{bif}
The difference between the bifurcation threshold obtained from the
finite-differences numerical simulation (left border of the red area in Fig.
\ref{FIG4}) and the full line obtained from a linear stability analysis of the
circular Couette flow is due to the limited numerical resolution of the
former. This difference increases with increasing $R_1, |R_2|$:  For higher
Reynolds numbers the vortex flow intensifies close to the inner cylinder. This
uneven flow distribution is better captured by the very high spatial
resolution used in the linear analysis than with the fixed homogeneous grids
used in the nonlinear finite-differences calculations. In Fig.~\ref{FIG4} the
differences do not exceed $3-4 \%$.
\end{thebibliography}
\end{document}